\documentclass[12pt]{article}

\usepackage{graphicx}
\usepackage[svgnames]{xcolor}
\usepackage{amsmath,amssymb}

\def\bea{\begin{eqnarray}}
\def\eea{\end{eqnarray}}
\def\bes{\begin{eqnarray}}
\def\ees{\end{eqnarray}}
\def\beq{\begin{eqnarray}}
\def\eeq{\end{eqnarray}}
\def\be{\begin{equation}}

\def\ee{\end{equation}}

\def\ba{\begin{array}}
\def\ea{\end{array}}
\def\bi{\begin{itemize}}
\def\ei{\end{itemize}}

\def\pK{{\color{Purple} K}}

\begin{document}
\title{Time-dependent flow from an AdS Schwarzschild black hole\footnote{Research supported in part by the US Department of Energy under grant DE-FG05-91ER40627.}}
\author{George Siopsis}
\author{George Siopsis\footnote{E-mail: siopsis@tennessee.edu}\\
\em Department of Physics
and Astronomy, \\
\em The University of Tennessee, Knoxville, \\
\em TN 37996 - 1200, USA.
}
\date{}

\maketitle

\vspace{-3.5in}\hfill UTHET-10-0301\vspace{3.5in}

\begin{abstract}
I discuss two examples of time-dependent flow which can be described in terms of
an AdS Schwarzschild black hole via holography. The first example involves Bjorken hydrodynamics which should be applicable to the formation of the quark gluon plasma in heavy ion collisions.
The second example is the cosmological evolution of our Universe.
\end{abstract}

\vspace{2in}

\noindent{\sl Prepared for the proceedings of the 6th International Symposium
on Quantum Theory and Symmetries, University of Kentucky, Lexington, KY, July 2009.}

\newpage

\section{Introduction}

The AdS/CFT correspondence affords us a means of studying the behavior of gauge theory fluids at strong coupling
in terms of gravity duals in a space of one extra dimension.
This is done by mapping properties of solutions of the Einstein field equations with a negative cosmological constant onto
the conformal boundary via holographic renormalization \cite{Skenderis,Skenderis2}.
Understanding the transport properties of strongly coupled time-dependent fluids is a challenge due to the lack of exact time-dependent supergravity solutions.
Here I discuss how this problem can be circumvented by applying holographic
renormalization to static AdS Schwarzschild black hole solutions.
I discuss two examples: Bjorken (boost-invariant) hydrodynamics which should be related to the quark-gluon plasma produced in heavy ion colliders (RHIC and the LHC) and the cosmological evolution of the Universe.

\section{AdS$_d$ Schwarzschild black holes}

AdS Schwarzschild black holes are exact solutions of the Einstein field equations in the presence of a negative cosmological constant.
The metric can be written as
\be\label{eqmebh} ds_{\mathrm{b.h.}}^2 = \frac{R^2}{z^2} \left( -f(z) dt^2 + d\Sigma_{\pK,d-2}^2 + \frac{dz^2}{f(z)} \right) \ \ , \ \ \ \ f(z) = 1+ \pK \frac{z^2}{R^2} -2\mu z^{d-1} \ee
I shall choose units so that the AdS radius $R=1$.
The
horizon radius satisfies $f(z_+)=0$ and the Hawking temperature is
\be\label{eq2}
T_H=\frac{d-1+\pK(d-3)z_+^2}{4\pi z_+}
\ee
The mass and entropy of the hole are, respectively,
\be\label{BH}
M=\frac{(d-2)(1+ \pK z_+^2)}{16\pi G z_+^{d-1}} Vol(\Sigma_{\pK,d-2})~,~~~ S=\frac{1}{4G z_+^{d-2}} Vol(\Sigma_{\pK,d-2})
\ee
The parameter $\pK$ determines the curvature of the horizon and the boundary of AdS space.
For
$\pK =0, +1, -1$ we have, respectively,
a flat ($\mathbb{R}^{d-2}$),
spherical ($\mathbb{S}^{d-2}$) and
hyperbolic ($\mathbb{H}^{d-2} /\Gamma$, topological black hole, where
$\Gamma$ is a discrete group of isometries)
horizon (boundary).


For simplicity, I shall concentrate on a flat
($\pK = 0$) hole which is the large hole limit of any AdS Schwarzschild black hole.
It has a flat horizon at
$z_+ = (2\mu)^{-\frac{1}{d-1}}$.
This black hole is dual to a gauge theory plasma residing on the conformal boundary
via 
holographic renormalization \cite{Skenderis,Skenderis2}.
To see this, write the metric (\ref{eqmebh}) in terms of
Fefferman-Graham coordinates
\be\label{FG}
ds^2=\left( g_{\mu\nu}dx^\mu dx^\nu+dz_{FG}^2 \right) / z_{FG}^2\ee
Near the
boundary ($z_{FG}=0$), expand
\be
g_{\mu\nu} = g_{\mu\nu}^{(0)}+z_{FG}^2 g_{\mu\nu}^{(2)} +
\dots+z_{FG}^{d-1}g_{\mu\nu}^{(d-1)} +h_{\mu\nu}^{(d-1)}z_{FG}^{d-1}\ln z_{FG}^2 +\mathcal{O}(z_{FG}^d) \ee
where $g^{(0)}_{\mu\nu}=\eta_{\mu\nu}$.
The vacuum expectation value of the
stress-energy tensor of the plasma on the boundary is related to the bulk metric through
\be\label{eqgT} \langle T_{\mu\nu}\rangle = \frac{d-1}{16\pi G} g_{\mu\nu}^{(d-1)}
\ee
From this, we read off the energy density and pressure, respectively,
\be \epsilon = \langle T^{tt} \rangle = (d-2) \frac{\mu}{8\pi G} \ \ , \ \ \ \ p = \langle T^{ii} \rangle = \frac{\mu}{8\pi G} \ee
obeying $p = \frac{1}{d-2} \epsilon$, as expected for a conformal fluid.
We deduce the
equation of state
\be p = \frac{1}{16\pi G_d} \left( \frac{4\pi T_H}{d-1} \right)^{d-1} \ee
and energy and entropy densities, respectively, as functions of temperature
\be\label{eqes} \epsilon = \frac{d-2}{16\pi G} \left( \frac{4\pi T_H}{d-1} \right)^{d-1} \ \ , \ \ s = \frac{dp}{dT} = \frac{1}{4G} \left( \frac{4\pi T_H}{d-1} \right)^{d-2} \ee
Thus we obtain a static fluid at temperature $T=T_H$ (eq.~(\ref{eq2})).

\section{Bjorken flow}

Following a suggestion by Bjorken \cite{Bjorken}, in order to understand the
quark gluon plasma produced in heavy ion collisions, one needs to study boost
invariant hydrodynamics.
This is because there is a
``plateau'' in particle production in the central rapidity region.
The emergent nuclei are highly Lorentz-contracted pancakes
receding at speed of light,
resulting in a flow which is
independent of the Lorentz frame (boost invariant).
Given appropriate initial conditions (see table \ref{table1} for RHIC and the LHC),
\begin{table}[h]
\caption{\label{table1}Initial conditions for the quark gluon plasma at RHIC and the LHC}
\begin{center}
\begin{tabular}{||l||c|c|c|c||}
\hline\hline
 & $\tau_0$ (fm/c) & $\epsilon_0$ (GeV/fm$^3$) & $T$ (GeV) & $\sqrt s$ (GeV) \\
\hline
\hline
RHIC &  0.2 & 10 & 0.5 & 200 \\
\hline
LHC &  0.1 & 10 &  1 & 5,500 \\
\hline
\hline
\end{tabular}
\end{center}
\end{table}
the hydrodynamic equations
respect their symmetry (boost invariance)
and lead to
simple solutions.
For a conformal fluid, one obtains
\be \epsilon \sim \tau^{-4/3} \ \ , \ \ \ \ T \sim \tau^{-1/3} \ \ , \ \ \ \ s \sim \tau^{-1} \ee
where $\tau$ is the longitudinal proper time defined in (\ref{eq11}).

Suppose that the plasma lives on a $(d-1)$-dimensional Minkowski space spanned by
coordinates $\tilde x^\mu$ ($\mu = 0,1,\dots, d-2$)
and the
colliding beams are along the $\tilde x^1$ direction.
Introducing coordinates $\tau, y$ (proper time and rapidity in the longitudinal plane, respectively)
\be\label{eq11}
\tilde x^0=\tau \cosh y~~,~~~~\tilde x^1=\tau \sinh y
\ee
the Minkowski metric reads
\be\label{eq16}
d\tilde s^2=d\tilde x_\mu d\tilde x^\mu = -d\tau^2+\tau^2 dy^2+(d\tilde x^\bot)^2
\ee
where $\tilde x^\bot = (\tilde x^2,\dots,\tilde x^{d-2})$ are the transverse coordinates.

The stress-energy tensor is
\be
T^{\mu\nu} =
\mathrm{diag} \left[
\epsilon (\tau) \ \ \ \
\frac{p(\tau)}{\tau^2} - 2 \frac{d-3}{d-2} \frac{\eta(\tau)}{\tau^3} \ \ \ \
\dots \ \ \ \
p(\tau)+ \frac{2}{d-2} \frac{\eta(\tau)}{\tau}
\right] \ .
\ee
The local conservation law yields
\be
\partial_\tau \epsilon +\frac{1}{\tau} (\epsilon + p)  - 2\frac{d-3}{d-2} \frac{\eta}{\tau^2} =0
\ee
Conformal symmetry implies tracelessness ($T_\mu^\mu =0$), therefore
\be\label{HydroEP}
\epsilon = (d-2) p = \epsilon_0\tau^{-\frac{d-1}{d-2}}
\ee
Then conservation of entropy in perfect fluid implies
\be\label{HydroT}
T=\frac{T_0}{\tau^{1/(d-2)}}
\ \ , \ \ \
s = \frac{\dot p}{\dot T} = \frac{s_0}{\tau} \ \ , \ \ \ \ s_0 = \frac{d-1}{d-2}\, \frac{\epsilon_0}{T_0} \ee
Notice that the energy and entropy densities have the same dependence on temperature as in the static case (eq.~(\ref{eqes})).
If we identify initial data (at $\tau =1$) with their corresponding values in the static case,
\be\label{eqTe} T_0 = T_H \ \ , \ \ \ \ \epsilon_0 = \frac{(d-2)\mu}{8\pi G} \ee
then eq.~(\ref{eqes}), with $T_H$ replaced by $T$ (\ref{HydroT}), describes the evolution of the energy and entropy densities in a Bjorken flow.

One may similarly find the time dependence of viscosity corrections.
At high temperatures one expects that the ratio $\eta /s$ asymptotes to a constant.
Therefore,
\[ \eta = \eta_0\tau^{-1} + \dots \ , \ \
\epsilon = \epsilon_0\tau^{-\frac{d-1}{d-2}} - 2\eta_0\tau^{-2} + \dots \]
\be T = T_0 \left[ \frac{1}{\tau^{\frac{1}{d-2}}} - \frac{2\eta_0}{(d-1)\epsilon_0\tau} + \dots \right] \ , \ \
s = s_0 \left[ \frac{1}{\tau} - \frac{2(d-2)\eta_0}{(d-1)\epsilon_0} \frac{1}{\tau^{\frac{2d-5}{d-2}}} + \dots \right] \ee
For the gravity dual of Bjorken flow, a
solution of the Einstein equations
was constructed
\cite{Janik,Janik2,Nakamura}
as a series expansion in large longitudinal time keeping
the parameter
\be v = \tilde z\tau^{-1/(d-2)} \ee
fixed.
At leading order the metric that leads to boost invariant hydrodynamics reads
\be\label{metric}
ds^2_{\mathrm{Bjorken}} =\frac{1}{\tilde z^2} \left[ -\left( 1 -2\mu v^{d-1} \right) d\tau^2+\tau^2 dy^2+(d\tilde x^\bot)^2
+\frac{d\tilde z^2}{ 1 -2\mu v^{d-1} }  \right]
\ee
Higher-order corrections can be systematically calculated and lead to the ratio
\be\label{eq44} \frac{\eta}{s} = \frac{1}{4\pi} \ee
as with sinusoidal perturbations of black holes \cite{Policastro}.


Interestingly, the above asymptotic expression (\ref{metric}) for the metric can also be obtained from a static Schwarzschild black hole as follows \cite{bib1a}.
Instead of
approaching the conformal boundary with $z=$~const.~hypersurfaces (as $z\to 0$), use
$\tilde z=$~const.~hypersurfaces, where
\be\label{eq23}
t=\frac{d-2}{d-3} \tau^{\frac{d-3}{d-2}}~~,~~~~x^1=\tau^{\frac{d-3}{d-2}} y
~~,~~~~x^\bot=\frac{\tilde x^\bot}{\tau^{1/(d-2)}}~~,~~~~z=\frac{\tilde z}{\tau^{1/(d-2)}}
\ee
They coincide initially (at $\tau = 1$) with their static counterparts, but they
``flow'' because the new coordinates (\ref{eq23}) of the black hole metric are $\tau$-dependent.

Applying the transformation (\ref{eq23}) to the exact black hole metric (\ref{eqmebh}), or
more precisely, to a patch which includes the boundary $z\to 0$,
one obtains
\be ds^2_{\mathrm{b.h.}} = \frac{1}{\tilde z^2} \left[ - \left( 1-2\mu v^{d-1} \right) d\tau^2 + \tau^2 dy^2 + (d\tilde x^\bot)^2
+ \frac{d\tilde z^2}{1-2\mu v^{d-1}} \right] + \dots
\ee
which matches the bulk metric of Bjorken flow (\ref{metric}) to leading order in $1/\tau$ with fixed $v$.
Thus, one obtains Bjorken hydrodynamics asymptotically (as $\tau\to\infty$) from a static black hole!

Higher-order corrections to Bjorken flow dictated by the black hole may be found by refining the transformation \cite{bib1b}.
This entails introducing corrections which are of $o(1/\tau)$ and making sure that the application of the transformation to the black hole metric does not introduce dependence of the metric on the rapidity and the transverse coordinates.
It can be done systematically at each order in the $1/\tau$ expansion.
However, at higher orders, to maintain boost invariance on the boundary, one must perturb the Schwarzschild metric.
This perturbation leads to the ratio (\ref{eq44}) \cite{bib1b}
in agreement with the time-dependent asymptotic solution of Einstein equations.

\section{Cosmology}


In a popular scenario for cosmology, our Universe is modeled by a 3-brane 
embedded in a higher-dimensional bulk.
The cosmological evolution of the brane  
is equivalent to its
motion within the bulk space. 
The bulk may be occupied by a black hole, or a more complicated (or unknown) 
solution to the Einstein equations
(allowing for energy exchange between the brane and
the bulk).

My goal is to
understand how the cosmological evolution emerges in the context of 
the AdS/CFT correspondence.
I shall show that the equations of cosmological evolution emerge via
holographic renormalization, even when one starts
from a static AdS Schwarzschild black hole, provided the boundary conditions are chosen 
appropriately \cite{bib1c}.
I shall concentrate on the case of a five-dimensional bulk $\mathcal{M}$ with a physically relevant four-dimensional conformal boundary $\partial\mathcal{M}$.

One usually fixes the geometry of the conformal boundary by adopting Dirichlet boundary conditions.
However,
for cosmological evolution 
the boundary geometry must remain dynamical.
This may lead to fluctuations of the bulk metric which are not normalizable,
but such fears are unfounded: the boundary geometry can 
be dynamical if one correctly introduces boundary counterterms needed in order to 
cancel infinities \cite{Compere}.

The Einstein field equations are obtained by varying the bulk action 
$I_{\mathcal{M}}$ which consists of the
five-dimensional Einstein-Hilbert action on
$\mathcal{M}$ with a cosmological term,
the Gibbons-Hawking boundary 
term and
boundary counterterms needed to render the system finite.

In addition to a black hole solution in the bulk,
one thus obtains the boundary term
\be\label{eqvar} \delta I_{\mathcal{M}} = \frac{1}{2} \int_{\partial \mathcal{M}} 
d^4 x \sqrt{-\det g^{(0)}} T_{\mu\nu}^{(CFT)} \delta g^{(0)\mu\nu} \ee
where $g_{\mu\nu}^{(0)}$ ($\mu,\nu = 0,1,2,3$) is the boundary metric
and $T_{\mu\nu}^{(CFT)}$ is the stress-energy tensor of the dual
CFT on the conformal boundary $\partial\mathcal{M}$.
Dirichlet boundary conditions fix $g_{\mu\nu}^{(0)}$, 
therefore the additional term (\ref{eqvar}) vanishes.

To keep $g_{\mu\nu}^{(0)}$ dynamical,
I shall adopt {\em mixed} boundary conditions.
To this end,
introduce the boundary action
\be\label{eq5a} I_{\partial\mathcal{M}} = 
I_{\partial\mathcal{M}}^{(EH)} + I_{\partial\mathcal{M}}^{(matter)} \ . \ee
consisting of the four-dimensional Einstein-Hilbert action
\be I_{\partial\mathcal{M}}^{( EH)} = - \frac{1}{16\pi G_4} 
\int_{\partial\mathcal{M}} d^4 x \sqrt{-\det g^{(0)}} (\mathcal{R} - 2\Lambda_4) \ , 
\ee
where $\quad G_4$ ($\Lambda_4$) is Newton's (cosmological) constant in 
the four-dimensional boundary
and
$\quad \mathcal{R}$ is the four-dimensional
Ricci scalar constructed from $g_{\mu\nu}^{(0)}$,
and an
unspecified action for matter fields,
\be\label{eqmat4} I_{\partial\mathcal{M}}^{( matter)} = 
\int_{\partial \mathcal{M}} d^4 x \sqrt{-\det g^{(0)}} \mathcal{L}^{(matter)} \ . \ee
which may reside on the boundary and have no bulk duals.

To the variation of the bulk action
one must now add the variation of the new boundary action 
$\delta I_{\partial\mathcal{M}}$
under a change in the boundary metric.
\be \delta I_{\mathcal{M}} +\delta I_{\partial\mathcal{M}} =0 \ee
This leads to the possibility of
{\em mixed} boundary conditions,
\be\label{eqcosmo} \mathcal{R}_{\mu\nu} - \frac{1}{2} g_{\mu\nu}^{(0)} \mathcal{R} 
- \Lambda_4 g_{\mu\nu}^{(0)} = 8\pi G_4 \left( T_{\mu\nu}^{(CFT)} 
+ T_{\mu\nu}^{(matter)} \right) \ .\ee
which are simply the four-dimensional Einstein field equations!
Moreover, the variation of the boundary matter action under a
change in the matter fields yields
the standard four-dimensional matter field equations.

For the most straightforward application of the AdS/CFT correspondence, write the metric (\ref{eqmebh}) in 
Fefferman-Graham coordinates
\be \frac{dz_{FG}}{z_{FG}} = \frac{dz}{z\sqrt{f(z)}} \ee
which gives (with an appropriate integration constant)
\be z_{FG}^4 = \frac{16}{\pK^2+4\mu} \ \frac{1 + \frac{\pK}{2} z^2 - \sqrt{f(z)}}{1 
+ \frac{\pK}{2} z^2+ \sqrt{f(z)}} \ \ , \ \ \ \
z^2 = \frac{z_{FG}^2}{\alpha + \beta z_{FG}^2 + \gamma z_{FG}^4} \ee
where
$\alpha = 1$,
$\beta = - \frac{\pK}{2}$, $\gamma = \frac{\pK^2+4\mu}{16}$.
The metric (\ref{eqmebh}) reads
\be\label{eqmetric1} ds_{\mathrm{b.h.}}^2 = \frac{1}{z_{FG}^2} 
\left[ dz_{FG}^2 - \frac{\left( 1- \gamma z_{FG}^4 \right)^2}{1+\beta z_{FG}^2 + \gamma z_{FG}^4} dt^2 
+ \left( 1+ \beta z_{FG}^2 +\gamma z_{FG}^4 \right) d\Sigma_{\pK,3}^2 \right] \ee
Through
holographic renormalization \cite{Skenderis,Skenderis2}, we deduce
the energy density and pressure, respectively,
\be\label{eq3} \langle T_{tt}^{(CFT)} \rangle = 3\langle T_{ii}^{(CFT)} \rangle = 
\frac{3\gamma}{4\pi G_5}  \ee
on a static Einstein Universe $\mathbb{R}\times \Sigma_{\pK ,3}$ with metric
\be\label{eqmetric0} ds_0^2 = g_{\mu\nu}^{(0)} dx^\mu dx^\nu = -dt^2 + d\Sigma_{\pK,3}^2 \ee
It should be noted that the
total energy $E=\langle T_{tt}^{(CFT)} \rangle Vol(\Sigma_{\pK,3})$ is larger 
than the mass of the black hole by a constant (Casimir energy) in the 
case of a curved horizon ($\pK\ne 0$).
The two quantities agree for flat horizons ($\pK=0$). 
The additional piece is due to a change of the vacuum 
state from Minkowski to the conformal vacuum
(curved metrics 
can be conformally mapped on Minkowski space).


For cosmology,
instead of a static boundary, we need one with the 
form of a {\it Robertson-Walker} (RW) spacetime
\be\label{eqmetricb} ds_0^2 = g_{\mu\nu}^{(0)} dx^\mu dx^\nu = 
-d\tau^2 + a^2(\tau) d\Sigma_{\pK,3}^2 \ee
To this end, I shall choose a different foliation away from the black hole, 
consisting of hypersurfaces whose metric is asymptotically RW.
To implement this,
change coordinates
$(t,z) \to (\tau, z_{FG})$ and bring the black-hole metric
in the form
%
\be\label{eqmetric3} ds_{\mathrm{b.h.}}^2 = \frac{1}{z_{FG}^2} \left[ dz_{FG}^2 
- \mathcal{N}^2(\tau,z_{FG}) d\tau^2 + \mathcal{A}^2 (\tau,z_{FG}) d\Sigma_{\pK,3}^2 \right]\ , \ee
where $\mathcal{N}(\tau,z_{FG})\to 1$ and $\mathcal{A}(\tau,z_{FG})\to a(\tau)$ 
as we approach the boundary $z=0$.
Comparison with the static case suggests the {\em ansatz}
\be \mathcal{A}^2 = \alpha(\tau) + \beta(\tau) z_{FG}^2 + \gamma (\tau) z_{FG}^4 \ , \ee
where $\alpha(\tau), \beta(\tau), \gamma(\tau)$ are functions to be determined.

$\mathcal{N}$ is constrained by the $\tau z_{FG}$ component of the 
Einstein equations,
\be \mathcal{N} = \frac{\dot{\mathcal{A}}}{\delta (\tau)} \ . \ee
Agreement with the boundary metric fixes
\be \alpha (\tau) = a^2 (\tau) \ \ , \ \ \ \ \delta (\tau) = \dot a (\tau) \ . \ee
The diagonal components of the Einstein equations collectively yield
\be \beta = - \frac{\dot a^2 + \pK}{2} \ .\ee
The rest of the Einstein equations are satisfied provided
\be \beta \dot\beta = 2(\dot\alpha\gamma + \alpha\dot\gamma)\ , \ee
which is integrated to give
\be\label{eqgam} \gamma = \frac{(\dot a^2 +\pK)^2 +4\mu}{16a^2} \ , \ee
where we fixed the integration constant by comparing with the 
static case.
It follows that the
metric {\it ansatz} is uniquely specified. 
It
agrees with the black hole metric (\ref{eqmebh}) provided
\be\label{eqsys1} \frac{(z')^2}{f(z)} - f(z) (t')^2 = \frac{z^2}{z_{FG}^2} \ , \ \
\frac{z' \dot z}{f(z)} - f(z) t'\dot t = 0 \ , \ \
\frac{\dot z^2}{f(z)} - f(z) \dot t^2 = - \frac{z^2\mathcal{N}^2}{ z_{FG}^2} \ , \ \
1 = \frac{z\mathcal{A}}{z_{FG}} \ee
The last equation fixes $z(\tau,z_{FG})$.
Two of the other three equations
determine the derivatives $\dot t= \frac{\dot A z'}{f \dot a} $ and $t'= -\frac{z^2\dot a}{z_{FG} f} $.
These expressions actually satisfy all three equations.
One
can verify consistency of the system by calculating
the mixed derivative $\dot t'$ using each of the two expressions
and showing that they match. 
Upon integration, we obtain a unique function $t(\tau,r)$, up to an irrelevant constant.

As an example, consider
pure AdS space in Poincar\'e coordinates ($\mu = 0$, $\pK=0$). We have
\be t(\tau ,z_{FG}) = -\frac{2\dot a z_{FG}^2}{4a^2 - {\dot a^2} z_{FG}^2} 
+ \int^\tau \frac{d\tau'}{a(\tau')} \ee
At the
boundary ($z_{FG}=0$), $t$ reduces to conformal 
time, $\int^\tau {d\tau'}/{a(\tau')}$.
It
receives corrections as we move into the bulk.
This is generally the case.

In particular, for a de Sitter (dS) boundary,
$a(\tau) = e^{H\tau} $, the metric reads
\be ds^2 = \frac{1}{z^2} \left[ dz^2 + \left( 1 - \frac{H^2}{4} z^2 \right) \left\{ -d\tau^2 + e^{2H\tau} d\vec{x}^2 \right\} \right] \ee
There is an {\it apparent horizon} at
$z_+ = \frac{2}{H}$ and the
temperature (from the conical singularity
of Euclidean space) is
\be\label{eqTdS} T = \frac{H}{2\pi} \ee
which coincides with the temperature of four-dimensional dS space
in the conformal vacuum!

General explicit expressions are not needed to extract 
physical results
since we already know the explicit form of the metric 
in the new coordinates.
The
energy density and pressure, respectively, are
\be\label{eq4} \langle ( T^{(CFT)} )_{\tau \tau} \rangle 
= \frac{3}{64\pi G} \ \frac{(\dot a^2+\pK)^2 + 4\mu}{a^4} 
\ , \ \
\langle ( T^{(CFT)} )_{i}^i \rangle =  \frac{(\dot a^2 + \pK)^2+4\mu 
-4a\ddot a (\dot a^2 + \pK)}{64\pi Ga^4} \nonumber \ee
where there is no summation over $i$
(it can be chosen in any 
spatial direction due to isotropy).

We deduce the
conformal anomaly
\be g^{(0)\mu\nu} \langle T_{\mu\nu}^{(CFT)} \rangle = 
- \frac{3\ddot a (\dot a^2 + \pK)}{16\pi Ga^3} \ee

It should be pointed out that the above results can also be obtained by a
direct (four-dimensional) calculation
by
observing that the RW metric
is conformally equivalent to the flat Minkowski metric.

The temperature on the RW boundary is found to be
\be\label{eqTRW} T = \frac{T_H}{a} \ee
where $T_H$ is the Hawking temperature (temperature of a static boundary)
and $a$ is the conformal factor.
The entropy is the Bekenstein-Hawking entropy of the hole (\ref{BH})
obeying the law of thermodynamics
$dE = TdS-pdV$.

The boundary conditions yield the equation of cosmological 
evolution
\be H^2 + \frac{\pK}{a^2} -\frac{\Lambda_4}{3}
= \frac{1}{16\pi G} \left[ \left(  H^2 + \frac{\pK}{a^2} 
\right)^2 + \frac{4\mu}{a^4} \right] + \frac{8\pi G_4}{3} \rho \ee
where $\rho = T_{00}^{(matter)}$ is the energy density of four-dimensional
``ordinary'' matter (without a bulk dual)
and
$H = \frac{\dot a}{a}$ is the Hubble parameter.
This equation is of the expected form \cite{Kiritsis}, reflecting the
conformal anomaly and the presence of a radiative energy component with energy
density $\sim a^{-4}$.

\begin{figure}[ht]
\includegraphics[width=12pc]{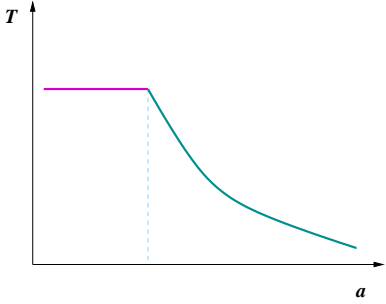}\hspace{2pc}%
\begin{minipage}[b]{18pc}\caption{\label{fig1}Evolution of the Universe.}
\end{minipage}
\end{figure}
To end on an amusing note, consider a phase transition (depicted in fig.~\ref{fig1}) between an exponentially expanding (dS) Universe whose bulk dual contains no black hole ($\mu =0$) at constant temperature (\ref{eqTdS}) and our Universe of decreasing temperature (\ref{eqTRW}) which is dual to a Schwarzschild black hole, as we discussed.
The transition occurs when the black hole forms. Did the formation of a black hole in the bulk cause the end of inflation?

%
%

\section{Conclusion}

Black holes and their perturbations are a powerful tool in understanding the hydrodynamic behavior of a gauge theory fluid at strong coupling.
Bjorken flow can be described in terms of a dual black hole in the bulk.
RHIC and the LHC may probe black holes and provide information on string theory
as well as non-perturbative QCD effects.
The cosmological evolution may also be understood in terms of an AdS$_5$ Schwarzschild black hole.


\end{document}